\documentclass[prd,aps,showpacs,preprintnumbers,floatfix,nofootinbib,twocolumn]{revtex4-1}
\pdfoutput=1

\usepackage{mathrsfs,amsmath,amsthm,amssymb,color,epstopdf,verbatim,enumerate,graphicx}

\usepackage{float}
\usepackage{hyperref}
\hypersetup{colorlinks,linkcolor=cyan,citecolor=cyan,urlcolor =[rgb]{0.0,0.0,0.5}}
\usepackage[nameinlink,noabbrev]{cleveref}
\usepackage{ulem}
\usepackage{xcolor}

\begin{document}

\preprint{Published 
%as P.~He \& B.-Q.~Ma, 
in \href{https://doi.org/10.1140/epjc/s10052-024-12758-x}{Euro.Phys.J. C 84 (2024) 401}}

\title{Abnormal threshold behaviors of photo-pion production off the proton in the GZK region}
\author{Ping He$\,{}^{a}$}
\author{Bo-Qiang Ma$\,{}^{a,b,c}$}\email{mabq@pku.edu.cn}
\thanks{corresponding author.}

\affiliation{${}^a$School of Physics, Peking University, Beijing 100871, China\\
		${}^b$Center for High Energy Physics, Peking University, Beijing 100871, China\\
		${}^c$Collaborative Innovation Center of Quantum Matter, Beijing, China}
		
%\date{\today}

\begin{abstract}
The cosmic-ray spectrum structures help to study the acceleration and propagation mechanism of ultra-high energy cosmic rays, and these structures were predicted to culminate in a cut-oﬀ, named the Greisen-Zatsepin-Kuzmin (GZK) cut-off, near $5\times10^{19}~\mathrm{eV}$ as a result of the inelastic interaction of protons with the $2.73~\mathrm{K}$ black body radiation.
The confirmation of the existence of GZK cut-off was tortuous, leading to activities to explore new physics, such as the cosmic-ray new components, unidentified cosmic-ray origins, unknown propagation mechanism, and the modification of fundamental physics concepts like the tiny Lorentz invariance violation~(LV).
The confirmation of the GZK cut-off provides an opportunity to constrain the LV effect. 
We use a phenomenological framework to restudy the GZK mechanism  
under the Planck scale deformation of the proton and pion dispersion relations. 
Restudying the photon induced pion production of the proton $\mathrm{p}+\gamma\to\mathrm{p}+\pi^0$, we predict abnormal threshold behaviors of this reaction under different LV modifications. 
Therefore, we can study the LV effects not only from the conventional GZK cut-off, but also from potentially threshold anomalies of the pion production process.
We divide the LV parameter space into three regions, and analyze the constraints from current observations in each region. 
The current observations have set strict constraints on a certain LV region. 
However, for others LV regions, further experimental observations and theoretical researches are still needed, and we also find survival space for some theoretical explorations that permit specific LV effects.
\end{abstract}

\maketitle

%\begin{twocolumn}

\section{Introduction}

%宇宙射线光谱的发现和研究历史
Ultra-high energy~(UHE) cosmic rays are the highest energy particles ever observed by human being. 
The cosmic rays with energies above $10^{19}~\mathrm{eV}$ were first detected 60 years ago~\cite{P36-Linsley-1961-extremely,P37-Linsley-1963-evidence}.
To study the acceleration and propagation mechanism of UHE cosmic rays, a variety of experiments have studied the cosmic-ray spectrum at extremely high energy. 
%光谱的基本特征
The cosmic-ray spectrum appears at ﬁrst to be a simple power law, and this spectrum exhibits signiﬁcant structures which can be used to 
%reﬂect 
study the cosmic-ray origins and propagations. 
At energy about $10^{15}~\mathrm{eV}$, the spectrum departs from original power law and steepens with a break known as the ``knee''. 
Then the spectrum occurs a second ``knee'' near $3\times10^{17}~\mathrm{eV}$, and above $10^{18}~\mathrm{eV}$ the spectrum has a dip called ``ankle''.
%GZK的提出
All these structures were predicted to culminate in a cut-oﬀ near $5\times10^{19}~\mathrm{eV}$, beyond which the spectrum drops abruptly. 
This cut-off was predicted in 1966 by K. Greisen, G. Zatsepin and V. Kuzmin~\cite{P1-Greisen-1966-end,P2-Zatsepin-1966-upper} as a result of the inelastic interaction of protons with the $2.73~\mathrm{K}$ black body radiation. 
Protons with energies above $5\times10^{19}~\mathrm{eV}$ could interact inelastically with the cosmic microwave background~(CMB) photons, producing pions and secondary hadrons with lower energies. 
Integrated over all possible sources in the Universe, it would produce a well-deﬁned break, dubbed the Greisen-Zatsepin-Kuzmin (GZK) cut-oﬀ.\\

%GZK的观测历史
Once, the confirmation of the existence of the GZK cut-oﬀ was a bumpy process as many initial attempts to detect this spectral feature did not ﬁnd it. 
%早期地面阵列没有能量观测GZK区域
The pioneering phase, exemplified by experiments such as Akeno, Haverah Park, Yakutsk and Fly’s Eye, had just barely enough sensitivity to begin to explore the GZK region~(for a review see Ref.~\cite{P3-Sokolsky-2007-Highest}).
%发现超出GZK-cutoff的光子
The Fly’s Eye air ﬂuorescence experiments measured an extraordinarily energetic event at $3\times10^{20}~\mathrm{eV}$~\cite{P4-Bird-1994-detection}, which attracted a great deal of attention. 
It was followed by the Akeno Giant Air Shower Array~(AGASA), which claimed that the cosmic-ray spectrum extends beyond $10^{20}~\mathrm{eV}$, and AGASA claimed that they saw more post-GZK events without candidate sources~\cite{P5-Takeda-1998-extension}. 
Only nearby sources~(closer than about $50~\mathrm{Mpc}$) could produce protons which would escape the GZK mechanism, due to lack of interaction length. 
However, none of the post-GZK events were pointed to any known active galaxy in our neighborhood.
The apparent lack of suitable astrophysical sources for these observed UHE cosmic rays is the ``GZK paradox''.  \\

%超出GZK-cutoff的数据的解释方法
Many exotic theoretical ideas were proposed to explain this apparent ``paradox".
Some hypotheses suggested new components of cosmic rays, such as magnetic monopoles~\cite{P13-Kephart-1995-magnrtic}, ``Z-boson bursts''~\cite{P14-Weiler-1997-cosmic}, and decay products of hypothetical super-heavy relic particles~\cite{P15-Berezinsky-1997-ultrahigh}.
Some theories proposed that these cosmic rays above the GZK cut-off come from unidentified cosmic-ray origins. 
Some theories explained that these post-GZK events originate from the propagation effect, such that these cosmic rays experience a large deflection under the extragalactic magnetic field structure~\cite{P12-Farrar-1999-GZK}.
%有可能来自于LV效应
There are also possibilities that these post-GZK events could be signals of Lorentz invariance violation~(LV)~\cite{P20-Coleman-1998-high,P21-Coleman-1998-evading,P22-Amelino-2000-planck,P23-Amelino-2001-space,P25-Gonzalez-1997-vacuum,P26-Sato-2000-extremely,P27-Bertolami-1999-proposed,P28-Aloisio-2000-probing}. 
Tiny departures from Lorentz invariance might have effects that rapidly increase with energy and kinematically prevent cosmic-ray nuclei from undergoing inelastic collisions with CMB photons.
%Coleman和Glashow 的观点
Coleman and Glashow proposed a perturbative framework, in which the tiny noninvariant terms were introduced into the standard model Lagrangian~\cite{P20-Coleman-1998-high,P21-Coleman-1998-evading}. 
These possible violations from strict Lorentz invariance can suppress or forbid inelastic collisions of cosmic-ray nuclei with CMB photons, and cause the GZK cut-off relaxed or removed.
%Giovanni Amelino-Camelia的观点
Giovanni Amelino-Camelia presented a general phenomenological framework for the description of Lorentz invariance deformation, and proposed that the LV effects could be induced by the nontrivial short-distance structure of space-time~\cite{P22-Amelino-2000-planck,P23-Amelino-2001-space}. 
Amelino-Camelia showed that a Planck scale deformation of the relativistic dispersion relation can explain the observations of post-GZK events, and he obtained constraints on LV effects from observation data.\\

\begin{figure}[H]
    \centering
    \includegraphics[scale=0.6]{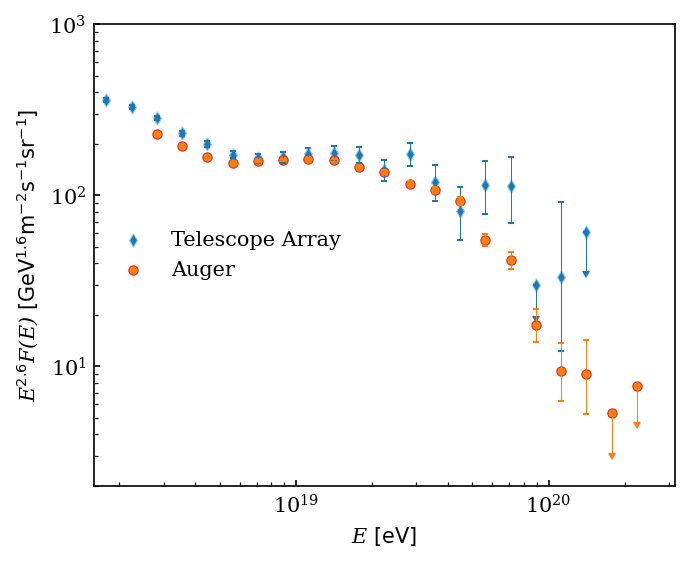}
    \caption{The highest-energy cosmic-ray spectrum from data of the Pierre Auger Observatory~\cite{P29-PierreAuger-2020-features,P30-PierreAuger-2020-measurement} and the Telescope Array~\cite{P35-Abbasi-2023-the}. The differential energy spectrum has been multiplied by $E^{2.6}$ in order to display the features of the steep spectrum that are otherwise difficult to discern.}
    \label{figgzkdata}
\end{figure}

%目前对GZK cutoff的确认
With the improvement of observation precision, the existence of GZK cut-off has been confirmed.
%HiRes确定了GZK-cutoff的存在
In 2004, the High Resolution Fly’s Eye~(HiRes) had produced data which clearly show the existence of a termination in the cosmic-ray ﬂux, consistent with the GZK cut-oﬀ prediction~\cite{P7-Thomson-2004-new,P8-HiRes-2004-measyrement,P9-HiRes-2005-observation,P10-HiRes-2008-first}. 
%近些年的观测数据及图像
With more and more observation equipments being put into use, the existence of GZK cut-off has been further confirmed.
%Pierre Auger Observatory的观测结果
Recently, Pierre Auger Observatory reported a measurement of the cosmic-ray spectrum above $2.5\times10^{18}~\mathrm{eV}$ based on $215030$ events~(from 2004/01/1 to 2018/08/31)~\cite{P29-PierreAuger-2020-features,P30-PierreAuger-2020-measurement}. 
Auger presented the steepening of the spectrum at around $5\times10^{19}~\mathrm{eV}$~\cite{P29-PierreAuger-2020-features,P30-PierreAuger-2020-measurement}. 
Auger presented that the spectral index changes from $2.51\pm0.03\mathrm{(stat)}\pm0.05\mathrm{(syst)}$ to $3.05\pm0.05\mathrm{(stat)}\pm0.10\mathrm{(syst)}$ at about $1.3\times10^{19}~\mathrm{eV}$ before increasing sharply to $5.1\pm0.3\mathrm{(stat)}\pm0.1\mathrm{(syst)}$ above $5\times10^{19}~\mathrm{eV}$~\cite{P29-PierreAuger-2020-features,P30-PierreAuger-2020-measurement}.
%Telescope Array的观测数据
Besides, Telescope Array reported the cosmic-ray spectrum observational data above $10^{17.5}~\mathrm{eV}$ from 2008/04/01 to 2017/11/28~\cite{P35-Abbasi-2023-the}. 
Through fitting of the spectrum to a series of broken power law models, Telescope Array indicated the GZK cut-off at $10^{19.8}~\mathrm{eV}$~\cite{P35-Abbasi-2023-the}.
Fig.~\ref{figgzkdata} shows the highest-energy cosmic-ray spectrum from data of the Pierre Auger Observatory~\cite{P29-PierreAuger-2020-features,P30-PierreAuger-2020-measurement} and the Telescope Array~\cite{P35-Abbasi-2023-the}. 
The observation of the GZK cut-off also set strong constraints on Lorentz-Violation parameters of the proton from theoretical studies~\cite{P38-Xiao-2008-lorentz,P39-Bi-2008-testing,P40-Stecker-2009-searching}. \\

%当前对GZK的确定就约束了LV效应
The confirmation of the GZK cut-off provides an opportunity for us to constrain the LV effect. 
We use a phenomenological framework to restudy the photon induced pion production of the proton $\mathrm{p}+\gamma\to\mathrm{p}+\pi^0$. 
Considering the Planck scale deformation of the dispersion relation, we predict abnormal threshold behaviors of the pion production under different LV modifications. Therefore in addition to the conventional GZK cut-off, we can also explore the LV effects of proton and pion from potentially threshold anomalies, in similar to the photon case discussed in the literature (see, e.g., Refs.~\cite{P41-Li-2021-threshold,P42-Li-2021-ultrahigh,P43-Li-2022-searching,P44-Li-2022-lorentz,P45-Li-2023-revisiting}).
We get the corresponding constraints on different proton and pion LV effects from the current observations, meanwhile, we also reveal survival space for some theoretical explorations that permit specific LV effects. \\

\section{Research and Discussion}

To review the GZK cut-off mechanism, and for convenience, we use the dispersion relations of proton, photon and pion to restudy the reaction $\mathrm{p}+\gamma\to\mathrm{p}+\pi^0$
\begin{equation}\label{equ1}
    \begin{cases}
    E^2_\mathrm{p}=p^2_\mathrm{p}+m^2_\mathrm{p} & \mathrm{proton};  \\
    \epsilon^2=q^2   &  \mathrm{photon};\\
    E^2_{\pi}=p^2_{\pi}+m^2_{\pi}  &  \mathrm{pion}. \\
    \end{cases}
\end{equation}
\begin{figure}[H]
    \centering
    \includegraphics[scale=0.6]{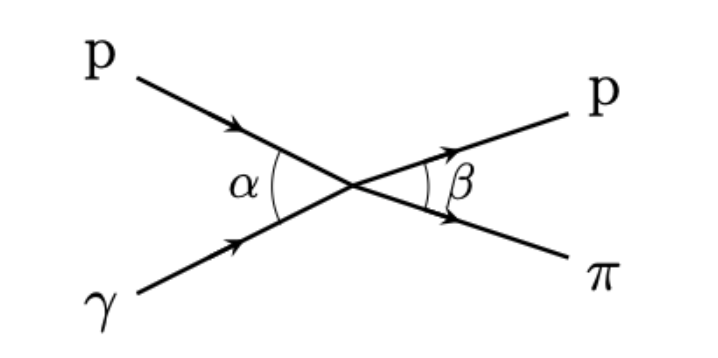}
    \caption{The diagram of the high-energy proton collision with a photon into a secondary proton and a pion.}
    \label{figdiagram}
\end{figure}

The diagram of $\mathrm{p}+\gamma\to\mathrm{p}+\pi^0$ is shown in Fig.~(\ref{figdiagram}), where the angle between the incoming proton and photon is $\alpha$ and the angle between the outgoing proton and pion is $\beta$. 
From special relativity, this reaction has a low threshold, which occurs when $\alpha=\pi$ and $\beta=0$. 
Because of this threshold occurrence condition, only considering the modulus of the momentum is reasonable. \\

A high-energy proton~$(E_\mathrm{p}, p_\mathrm{p})$ scatters with a photon~$(\epsilon, q)$ and produces a secondary proton~$(E'_\mathrm{p}, p'_\mathrm{p})$ and a pion~$(E_{\pi}, p_{\pi})$. The energy-momentum conservation relation of this reaction is
\begin{equation}\label{equ2}
    \begin{cases}
    E_\mathrm{p}(p_\mathrm{p})+\epsilon=E'_\mathrm{p}(p'_\mathrm{p})+E_{\pi}(p_{\pi});\\
    p_\mathrm{p}-q=p'_\mathrm{p}+p_{\pi}, \\
    \end{cases}
\end{equation}
and expanding to the leading-order of $(m/k)^2$, we get
\begin{equation}\label{equ3}
   p_\mathrm{p}+\frac{m_\mathrm{p}^2}{2p_\mathrm{p}}+q=p'_\mathrm{p}+\frac{m_\mathrm{p}^2}{2p'_\mathrm{p}}+p_{\pi}+\frac{m_{\pi}^2}{2p_{\pi}}.
\end{equation}
For a high-energy proton with energy $E_\mathrm{p}$, the minimum energy of the photon is~\footnote{The minimum photon energy is the reaction threshold of the photon. Studying the photon threshold behavior can constrain the LV effect of the GZK process. 
For example, under the LV assumption the photon energy threshold increases abruptly above an energy, and this phenomenon means that the LV effect will increase the GZK energy, resulting in UHE particles observable above the GZK region~\cite{P57-PierreAuger-2021-testing}.
In our work, we mainly study the threshold behavior of protons, which is beneficial for us to make better constraints from the spectrum structure.}
\begin{equation}\label{equ4}
   \epsilon_\mathrm{min}=\frac{1}{4E_\mathrm{p}}[(m_\mathrm{p}+m_{\pi})^2-m_\mathrm{p}^2].
\end{equation}
Correspondingly, for the CMB photon with energy $\epsilon$, the proton threshold is
\begin{equation}\label{equ5}
   E_\mathrm{p}=\frac{1}{4\epsilon}[(m_\mathrm{p}+m_{\pi})^2-m_\mathrm{p}^2].
\end{equation}
This threshold occurs where the outgoing particle energy-momentum distribution is
\begin{equation}\label{equ6}
    \begin{cases}
    E'_\mathrm{p}=\frac{m_\mathrm{p}}{m_\mathrm{p}+m_\pi} \cdot E_\mathrm{p};\\
    E_{\pi}=\frac{m_\pi}{m_\mathrm{p}+m_\pi} \cdot E_\mathrm{p}.\\
    \end{cases}
\end{equation}
%In classic case, this reaction has no upper threshold. 
For CMB photon, the target photon energies obey a thermal distribution with temperature $T=2.73~\mathrm{K}$, or $\omega_0\equiv kT=2.35\times10^{-4}~\mathrm{eV}$. 
If we simply consider the CMB characteristic energy as the photon reaction energy, the corresponding proton reaction threshold is $E_{\mathrm{p}, 0}=2.88\times10^{20}~\mathrm{eV}$. 
Above this threshold, protons should experience extensive collisions to lose energy, resulting in a sharp decrease in the spectrum. \\

To get the threshold behavior of the reaction $\mathrm{p}+\gamma\to\mathrm{p}+\pi^0$ under the LV effects, we modify the dispersion relations. 
The proton dispersion relation in LV case is
\begin{eqnarray}\label{equ7}
   E^2_\mathrm{p} &=& F(E_\mathrm{p}, p_\mathrm{p}; m_\mathrm{p}, \eta_\mathrm{p,n})  \notag \\ 
   &=& p^2_\mathrm{p}+m^2_\mathrm{p}-\eta_\mathrm{p,n}p^2_\mathrm{p}\left(\frac{E_\mathrm{p}}{E_\mathrm{Pl}}\right)^n,
\end{eqnarray}
and the pion dispersion relation in LV case is
\begin{eqnarray}\label{equ8}
   E^2_\pi&=& F(E_\pi, p_\pi; m_\pi, \eta_\mathrm{\pi, n})\notag\\ 
   &=&p^2_\pi+m^2_\pi-{\eta_\mathrm{\pi, n}}p^2_\pi \left(\frac{E_\pi}{E_\mathrm{Pl}}\right)^n,
\end{eqnarray}
where $E_\mathrm{Pl}\simeq1.22\times10^{19}~\mathrm{GeV}$ is the Plank scale. 
Since the LV effects are very tiny, we generally expect that LV effects could have observable effects in only extremely high energy region, and the LV modifications would be suppressed by the Planck scale~\footnote{Since the LV effects are very tiny, it is important to choose an appropriate method to characterise the LV effect. 
Some studies choose to measure the cosmic particle average free path in the CMB photons~\cite{P58-Maccione-2009-Planck}. 
When the LV effect lengthens the proton average free path, it means that the LV effect makes particles higher than GZK energy more easily observable. 
This method can more conveniently locate the origin of particles. 
In our work, we choose to use the proton threshold behavior to characterise the LV effect, which allows us to intuitively see the proton anomalous threshold performance in different LV parameters, thus obtaining a better understanding of the spectrum structure.}. 
$\eta_\mathrm{p, n}$ and $\eta_\mathrm{\pi, n}$ are the $n$th-order LV parameters of protons and pions respectively.
For convenience, if we only consider the linear modification, we set  $\eta_\mathrm{p, 1}\equiv\eta_\mathrm{p}$, $\eta_\mathrm{\pi, 1}\equiv\eta_\pi$.
Due to the low energy of background photons, we do not consider the modification of the photon dispersion relation.
For a threshold reaction, the threshold occurs when the final particle momenta are parallel and the initial momenta are antiparallel~\cite{P18-H6-Mattingly-2002-threshold}. 
So it is reasonable for us to only consider the modulus of the momentum $p=|\vec{p}|$.
Currently, there are many theories that can induce the violation of Lorentz symmetry~\cite{P19-H18-He-2022-lorentz}, and these theories may cause ambiguities when we study the LV effects. 
By choosing this phenomenological framework, we can study the proton reaction with photon under LV effects model-independently.
Next, we discuss different threshold behaviors under different LV modifications and the corresponding constraints on LV parameters from the observations.\\

\begin{widetext}
\subsection{Proton LV effect}

If we only consider the LV effect of the proton, we set $\eta_\mathrm{p}\equiv\eta$ for convenience, and the energy-momentum conservation relation Eq.~(\ref{equ2}) means that

\begin{equation}\label{equ9}
   p_\mathrm{p}+\frac{m_\mathrm{p}^2}{2p_\mathrm{p}}-\eta\frac{p_\mathrm{p}^2}{2E_\mathrm{Pl}}+q=p'_\mathrm{p}+\frac{m_\mathrm{p}^2}{2p'_\mathrm{p}}-\eta\frac{{p'}_\mathrm{p}^2}{2E_\mathrm{Pl}}+p_{\pi}+\frac{m_{\pi}^2}{2p_{\pi}}.
\end{equation}
We get:
\begin{equation}\label{equ10}
   \eta=\eta(E_\mathrm{p})=\frac{(m_\mathrm{p}+m_{\pi})^2}{(m_\mathrm{p}+m_\mathrm{\pi})^2-m_\mathrm{p}^2}\cdot f(E_\mathrm{p})=\frac{(m_\mathrm{p}+m_{\pi})^2}{(m_\mathrm{p}+m_\mathrm{\pi})^2-m_\mathrm{p}^2}\cdot4E_\mathrm{Pl}\cdot\left[\frac{\epsilon}{E_\mathrm{p}^2}-\frac{(m_\mathrm{p}+m_\mathrm{\pi})^2-m_\mathrm{p}^2}{4E_\mathrm{p}^3}\right],
\end{equation}
\end{widetext}
where we define a function $f(E_\mathrm{p})$. 
The proportion coefficient between $f(E_\mathrm{p})$ and $\eta(E_\mathrm{p})$ is $(m_\mathrm{p}+m_{\pi})^2/[(m_\mathrm{p}+m_\mathrm{\pi})^2-m_\mathrm{p}^2]$.
The threshold behaviors are equivalent to study the solutions of Eq.~(\ref{equ10}). 
Studying the properties of the function $f(E_\mathrm{p})$ helps us investigate the relation between $\eta$ and $E_\mathrm{p}$.
\begin{figure}[H]
    \centering
    \includegraphics[scale=0.6]{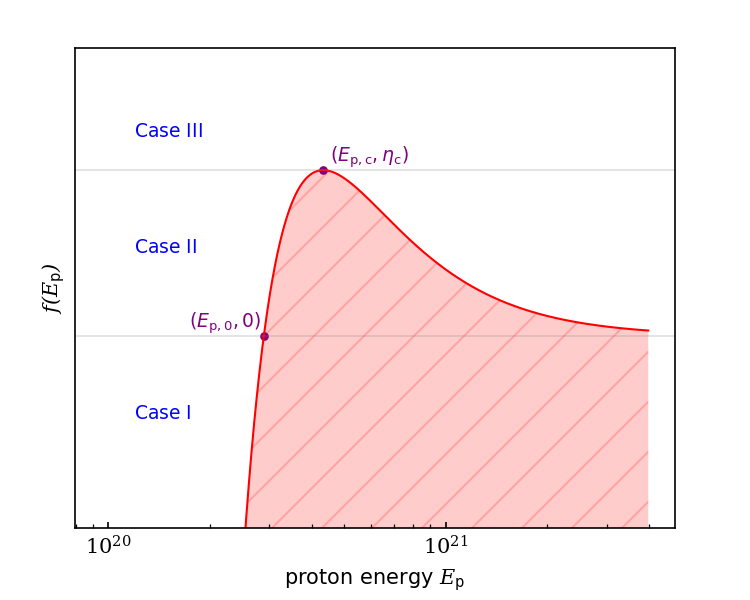}
    \caption{The trend diagram of $f(E_\mathrm{p})$. There are three reaction regions for proton LV parameter $\eta$ and proton energy $E_\mathrm{p}$.}
    \label{figetaenergyrelation}
\end{figure}

The image of $f(E_\mathrm{p})$ is represented schematically in Fig.~(\ref{figetaenergyrelation}), and the useful properties of $f(E_\mathrm{p})$ are as follows:
\begin{itemize}
    \item There is only one zero point $E_\mathrm{p,0}=[(m_\mathrm{p}+m_\mathrm{\pi})^2-m_\mathrm{p}^2]/4\epsilon$, which is the threshold derived in the classic case.
    \item The function tends to zero at $E_\mathrm{p}\to+\infty$, and tends to $-\infty$ at $E_\mathrm{p}\to0$.
    \item There is a maximum at the critical point $E_\mathrm{p,c}=\frac{3[(m_\mathrm{p}+m_\mathrm{\pi})^2-m_\mathrm{p}^2]}{8\epsilon}$  with the maximum $f_\mathrm{c}=\mathrm{max}f(E_\mathrm{p})=f(E_\mathrm{p,c})=\frac{256\epsilon^3E_\mathrm{Pl}}{27[(m_\mathrm{p}+m_\mathrm{\pi})^2-m_\mathrm{p}^2]^2}$.
\end{itemize}

%能量与破坏参数之间的关系
There are three regions in Fig.~(\ref{figetaenergyrelation}). 
Because $f(E_\mathrm{p})$ and $\eta(E_\mathrm{p})$ are linearly proportional, we will show that these three regions represent three different reaction threshold behaviors. 
The shade part is where the reaction occurs.
We get the properties of $\eta(E_\mathrm{p})$ from the properties of $f(E_\mathrm{p})$. 
Intersection point between shaded part and $\eta=0$ is $(E_\mathrm{p,0}, 0)$, where $E_\mathrm{p,0}=4\epsilon/[(m_\mathrm{p}+m_\mathrm{\pi})^2-m_\mathrm{p}^2]$ is the proton collision energy in the classic situation. 
The highest point of $\eta(E_\mathrm{p})$ is $(E_\mathrm{p,c},\eta_\mathrm{c})$, where $E_\mathrm{p,c}=\frac{3[(m_\mathrm{p}+m_\mathrm{\pi})^2-m_\mathrm{p}^2]}{8\epsilon}$ and $\eta_\mathrm{c}=\frac{(m_\mathrm{p}+m_{\pi})^2}{(m_\mathrm{p}+m_\mathrm{\pi})^2-m_\mathrm{p}^2}\cdot f_\mathrm{c}=\frac{(m_\mathrm{p}+m_{\pi})^2}{(m_\mathrm{p}+m_\mathrm{\pi})^2-m_\mathrm{p}^2}\cdot\frac{256\epsilon^3E_\mathrm{Pl}}{27[(m_\mathrm{p}+m_\mathrm{\pi})^2-m_\mathrm{p}^2]^2}$.
In these three regions, the threshold behaviors are different:
\begin{itemize}
    \item Case I, $\eta<0$. Similar to the classic situation, there is only a low threshold of the reaction $\mathrm{p}+\gamma\to\mathrm{p}+\pi^0$. Above this threshold, the reaction occurs and results in a drop on the cosmic-ray spectrum. Different from classic situation, the threshold varies with the proton LV parameter $\eta$. The proton LV effect is more obvious when $\eta$ is more far from $\eta=0$, i.e., the threshold is more low, and the subsequent cosmic-ray spectrum cut-off energy is more low.

    \item Case II, $0<\eta<\eta_\mathrm{c}$. In this situation, there are two thresholds $E_\mathrm{p,low}$ and $E_\mathrm{p, high}$. Above the low threshold $E_\mathrm{p,low}$, the reaction occurs and results in a drop on the cosmic-ray spectrum. But different from classic situation, there is an additional high threshold $E_\mathrm{p, high}$, above which the reaction does not occur. It renders cosmic rays above this high threshold can be observed. From the cosmic-ray spectrum, there is a cut-off at the low threshold and a reappearance at the high threshold. 

    \item Case III, $\eta>\eta_\mathrm{c}$. There is no threshold. In this situation, the proton LV effect is obvious enough, and the collision reaction does not occur. There is no cut-off can be observed in the cosmic-ray spectrum.
\end{itemize}

In Case I, $\eta<0$, and it corresponds to the proton superluminal LV effect. Because the threshold behavior in this region is similar to that in classic cases, with only a shift of the cut-off energy to a smaller value, it is difficult to detect its signal from the spectrum. 
This also means that this proton LV parameter has not been strictly excluded from the experiment, which may leave living space for some theories that allow for some specific proton LV effects.
If the future GZK region observation show that the GZK cut-off energy is smaller than the theoretical GZK cut-off value, it may be the signal of this LV effect,
and it would be very inspiring for the quantum gravity and grand unified theories, especially those still permit the existence of some specific LV modifications.\\

In Case II, there is an abnormal threshold behavior -- the reappearance of extremely high energy protons, that has been discussed for the first time for the proton case.
In Case II, $0<\eta<\eta_\mathrm{c}$, and it is a tiny proton subluminal LV effect. 
In this case, the spectrum breaks at the lower threshold and reappear above the upper threshold. 
If future observations reveal proton cosmic rays with energy higher than the GZK cut-off energy or even higher, it may be a signal of this LV effect.
It should be noted that Case III may also lead to the emergence of high-energy photons above the GZK-cutoff energy. 
It requires further observation to determine whether spectrum breaking at the lower threshold exists.
We should note that observing this phenomenon has a strict demand on the source: the astrophysical sources can accelerate protons to the corresponding energy range.\\

Case III corresponds to the strictly constrained area after the GZK structure is determined.
In Case III, the GZK cut-off observations set strict proton parameter constraint: $\eta<\eta_\mathrm{c}$.
If we consider the CMB characteristic energy $\omega_0\equiv kT=2.35\times10^{-4}~\mathrm{eV}$, we get corresponding constraint: $\eta<\eta_\mathrm{c}\simeq8.6\times10^{-17}$. 
This constraint on proton parameter means that the proton LV energy scale is 17 orders larger than the Planck scale.
The previous constraints obtained from Pierre Auger Observatory roughly correspond to the strict constraint $\eta<10^{-12\sim-10}$~\cite{P55-Lang-2020-Testing, P57-PierreAuger-2021-testing}, indicating that our constraint is very strict.\\

To sum up, there are strict constraints on LV effect in Case III from the current GZK region spectrum detection, but there is still space for the existence of the LV effect in Case I and further investigation is needed to testify the possible phenomenon predicted by the LV effect in Case II. From a theoretical perspective, it can be seen that although previous experimental observations have set strict constraints on the LV effect, it actually only means a strong constraint on the LV effect in Case III. For the Cases I and II, it still allows us to explore the quantum gravity and grand unified theory that allow for the LV effect, and this requires us to handle them more carefully and rigorously.\\

\begin{widetext}
\subsection{Proton and pion linear LV effects}\label{Proton and pion linear LV effects}

If we consider the linear LV effects of proton and pion, we use the linear LV parameters $\eta_\mathrm{p}$ and $\eta_\pi$.
The energy-momentum conservation relation Eq.~(\ref{equ2}) means that
\begin{equation}\label{equ11}
   p_\mathrm{p}+\frac{m_\mathrm{p}^2}{2p_\mathrm{p}}-\eta_\mathrm{p}\frac{p_\mathrm{p}^2}{2E_\mathrm{Pl}}+q=p'_\mathrm{p}+\frac{m_\mathrm{p}^2}{2p'_\mathrm{p}}-\eta_\mathrm{p}\frac{{p'}_\mathrm{p}^2}{2E_\mathrm{Pl}}+p_{\pi}+\frac{m_{\pi}^2}{2p_{\pi}}-\eta_{\pi}\frac{{p}_{\pi}^2}{2E_\mathrm{Pl}},
\end{equation}
and we get
\begin{equation}\label{equ12}
   \eta_\mathrm{p}\cdot\frac{(m_\mathrm{p}+m_\mathrm{\pi})^2-m_\mathrm{p}^2}{(m_\mathrm{p}+m_{\pi})^2}-\eta_{\pi}\cdot\frac{m_{\pi}^2}{(m_\mathrm{p}+m_{\pi})^2}=f(E_\mathrm{p})=4E_\mathrm{Pl}\cdot\left[\frac{\epsilon}{E_\mathrm{p}^2}-\frac{(m_\mathrm{p}+m_\mathrm{\pi})^2-m_\mathrm{p}^2}{4E_\mathrm{p}^3}\right].
\end{equation}
\end{widetext}
We notice that, Eq.~(\ref{equ12}) has the similar form with Eq.~(\ref{equ10}). 
The properties of $f(E_\mathrm{p})$ are also useful to study the properties of LV parameters of proton and pion. 
Different from only proton LV case, the proton and pion LV parameters are coupled together and have the similar relation with $E_\mathrm{p}$.\\

The constraints on proton and pion parameters from spectrum observation are still divided into three regions. 
In Case I, $\eta_\mathrm{p}\cdot\frac{(m_\mathrm{p}+m_\mathrm{\pi})^2-m_\mathrm{p}^2}{(m_\mathrm{p}+m_{\pi})^2}-\eta_{\pi}\cdot\frac{m_{\pi}^2}{(m_\mathrm{p}+m_{\pi})^2}<0$, and the threshold behavior in this region is similar to that in classic cases.
It is difficult to distinguish these LV effects from spectrum observations.
In Case II, $0<\eta_\mathrm{p}\cdot\frac{(m_\mathrm{p}+m_\mathrm{\pi})^2-m_\mathrm{p}^2}{(m_\mathrm{p}+m_{\pi})^2}-\eta_{\pi}\cdot\frac{m_{\pi}^2}{(m_\mathrm{p}+m_{\pi})^2}<f_\mathrm{c}$. 
If the proton and pion parameters are configured appropriately, we might observe the reappearance of extremely high energy photons above the upper threshold.
The spectrum might break at GZK cut-off and reappear at higher energy. 
In Case III, the confirmation of GZK cut-off means the strict constraint: $f(E_\mathrm{p})<f_\mathrm{c}$, that is to say, the corresponding parameter constraint space is $\eta_\mathrm{p}\cdot\frac{(m_\mathrm{p}+m_\mathrm{\pi})^2-m_\mathrm{p}^2}{(m_\mathrm{p}+m_{\pi})^2}-\eta_{\pi}\cdot\frac{m_{\pi}^2}{(m_\mathrm{p}+m_{\pi})^2}<f_\mathrm{c}$.
With the CMB characteristic energy $\omega_0\equiv kT=2.35\times10^{-4}~\mathrm{eV}$, the strict constraint set by the GZK structure is $0.24\cdot\eta_\mathrm{p}-0.02\cdot\eta_{\pi}<2.04\times10^{-17}$. 
In this constraint, the constraints of proton and pion modifications are combined, because we consider both linear modifications simultaneously~\footnote{By starting from the proton reaction threshold conditions, the proton and pion LV parameter constraints can also be obtained~\cite{P59-Jacobson-2002-Threshold}. 
The threshold conditions include the energy-momentum conservation relation and positive particle energy values.}.\\

If the proton and pion have the same LV effect~($\eta_\mathrm{p}=\eta_\pi$), the energy-momentum conservation relation Eq.~(\ref{equ2}) means that

\begin{equation}\label{equ13}
   \eta_\mathrm{p}\cdot\frac{2m_\mathrm{p}m_\pi}{(m_\mathrm{p}+m_{\pi})^2}=f(E_\mathrm{p}).
\end{equation}
Under this special assumption, the proton and pion LV parameter constraints are similar to the previous results. 
The parameter constraint space can also be divided into three regions. 
It is worth noting that the assumption, that the proton and pion LV parameters are the same, is a highly restrictive constraint, and we usually do not make such assumption.\\

\begin{widetext}
\subsection{Proton and pion $n$-order LV modification}

If we consider the $n$-order modification of the proton and pion dispersion relation
\begin{equation}\label{equ14}
    \begin{cases}
    E^2_\mathrm{p} = p^2_\mathrm{p}+m^2_\mathrm{p}-\eta_\mathrm{p,n}p^2_\mathrm{p}\left(\frac{E_\mathrm{p}}{E_\mathrm{Pl}}\right)^n;\\
    E^2_\pi = p^2_\pi+m^2_\pi-{\eta_\mathrm{\pi, n}}p^2_\pi \left(\frac{E_\pi}{E_\mathrm{Pl}}\right)^n, \\
    \end{cases}
\end{equation}
the energy-momentum conservation relation Eq.~(\ref{equ2}) means that
\begin{equation}\label{equ15}
   p_\mathrm{p}+\frac{m_\mathrm{p}^2}{2p_\mathrm{p}}-\eta_\mathrm{p, n}\frac{p_\mathrm{p}^{n+1}}{2E_\mathrm{Pl}^n}+q=p'_\mathrm{p}+\frac{m_\mathrm{p}^2}{2p'_\mathrm{p}}-\eta_\mathrm{p, n}\frac{{p'}_\mathrm{p}^{n+1}}{2E_\mathrm{Pl}^n}+p_{\pi}+\frac{m_{\pi}^2}{2p_{\pi}}-\eta_{\pi, \mathrm{n}}\frac{{p}_{\pi}^{n+1}}{2E_\mathrm{Pl}^n}.
\end{equation}
Then, we get:
\begin{equation}\label{equ16}
   \eta_\mathrm{p, n}\cdot\frac{(m_\mathrm{p}+m_\mathrm{\pi})^{n+1}-m_\mathrm{p}^{n+1}}{(m_\mathrm{p}+m_{\pi})^{n+1}}-\eta_{\pi, \mathrm{n}}\cdot\frac{m_{\pi}^{n+1}}{(m_\mathrm{p}+m_{\pi})^{n+1}}=f_\mathrm{n}(E_\mathrm{p})=4E_\mathrm{Pl}^n\cdot\left[\frac{\epsilon}{E_\mathrm{p}^{n+1}}-\frac{(m_\mathrm{p}+m_\mathrm{\pi})^2-m_\mathrm{p}^2}{4E_\mathrm{p}^{n+2}}\right],
\end{equation}
\end{widetext}
where we define a function $f_\mathrm{n}(E_\mathrm{p})$.
We notice that, when $n=1$, Eq.~(\ref{equ16}) turns to Eq.~(\ref{equ12}), and  $f_1(E_\mathrm{p})$ turns to $f(E_\mathrm{p})$. 
Studying the threshold behaviors in $n$-order modification is equal to study the solutions of Eq.~(\ref{equ16}). 
The properties of $f_\mathrm{n}(E_\mathrm{p})$ are useful to help us to investigate the constraints on $n$-order proton and pion LV parameters from the cosmic-ray spectrum observation. 
The properties of $f_\mathrm{n}(E_\mathrm{p})$ are similar to $f(E_\mathrm{p})$:
\begin{itemize}
    \item There is only one zero point $E_\mathrm{p,0}=[(m_\mathrm{p}+m_\mathrm{\pi})^2-m_\mathrm{p}^2]/4\epsilon$, which is the threshold derived in the classic case.
    \item The function tends to zero at $E_\mathrm{p}\to+\infty$, and tends to $-\infty$ at $E_\mathrm{p}\to0$.
    \item There is a maximum at the critical point $E_\mathrm{p,c}=\frac{n+2}{n+1}\cdot\frac{(m_\mathrm{p}+m_\mathrm{\pi})^2-m_\mathrm{p}^2}{4\epsilon}$ with $f_\mathrm{n, c}=\mathrm{max}f_\mathrm{n}(E_\mathrm{p})=f_\mathrm{n}(E_\mathrm{p,c})=\frac{4^{n+2}(n+1)^{n+1}}{(n+2)^{n+2}}\cdot\frac{\epsilon^{n+2}E_\mathrm{Pl}^n}{[(m_\mathrm{p}+m_\mathrm{\pi})^2-m_\mathrm{p}^2]^{n+1}}$.
\end{itemize}

Correspondingly, the constraint space on $n$-order proton and pion LV parameters are also divided into three regions. 
In Case I, $f_\mathrm{n}(E_\mathrm{p})<0$, and the threshold behavior is similar to the classic cases.
In Case II, $0<f_\mathrm{n}(E_\mathrm{p})<f_\mathrm{n, c}$. 
When there is an appropriate configuration of the $n$-order proton and pion parameters, we might observe the reappearance of extremely high energy protons. 
In Case III, the GZK cut-off means the strict constraint: $f_\mathrm{n}(E_\mathrm{p})<f_\mathrm{n, c}$, that is to say, the $n$-order proton and pion LV parameters are constrained strictly.
With $n=1$ and the CMB characteristic energy $\omega_0\equiv kT=2.35\times10^{-4}~\mathrm{eV}$, we get the linear constraints of proton and pion $0.24\cdot\eta_\mathrm{p}-0.02\cdot\eta_{\pi}<2.04\times10^{-17}$, which is same to the result in Section \ref{Proton and pion linear LV effects}.
When we consider $n=2$, we get corresponding constraint: $0.33\cdot\eta_\mathrm{p,2}-0.002\cdot\eta_{\pi, 2}<6.13\times10^{-10}$, which is the quadratic constraints on proton and pion from the GZK structure.
We can derive constraints for higher-order modifications. 
According to the expression of $f_\mathrm{n, c}$, we find that $f_\mathrm{n, c}$ will increase with the increase of $n$, which means that the constraints will weaken under higher-order.
For example, for $n=4$, $f_\mathrm{4, c}=6.94\times10^{5}$, which means a very broad constraints on proton and pion, and we need to obtain $4$-order constraints from other studies.
According to Eq.~(\ref{equ16}), considering the coefficients before the proton and pion LV parameters $\eta_\mathrm{p, n}$ and $\eta_\mathrm{\pi, n}$, we find that the coefficient before $\eta_\mathrm{p, n}$ increases as $n$ increases, but the corresponding coefficient before $\eta_\mathrm{\pi, n}$ decreases. 
This means that for higher-order modifications, the constraint on the proton LV parameter $\eta_\mathrm{p, n}$ is stronger compared to that on the pion parameter $\eta_\mathrm{\pi, n}$.\\

Our research has an important assumption: the proton pion producation process dominates the structural changes in the cosmic-ray GZK region. This assumption is influenced by two factors: the acceleration mechanism of the astrophysical sources and the energy loss during propagation.
%GZK区域的宇宙射线被认为起源于河外
It is generally believed that the cosmic rays above the ``ankle" originate outside the Milky Way, and the acceleration process at the sources gives the initial values of the cosmic-ray components.
%加速机制的一般规律
Generally, the acceleration efficiency is proportional to the nucleus charge, i.e., the heavier elements are boosted to higher energies than light elements, but all these elements are limited by some maximum energy.
%我们的假设要求
Our hypothesis requires that the astrophysical sources can accelerate protons beyond the GZK energy range, which has a weak requirement for the maximum velocity of protons at the source.
This means that the proton spectrum changes in the GZK region is a propagation effect.
%源处的组分
At the same time, the requirement for protons to dominate means that the main high-energy particles at the source are protons, or other atomic nuclei will be disappeared through their propagation in the Universe by the interaction with the background lights.\\

Since the ``ankle" is traditionally thought of as the energy region where extragalactic origins begin to dominate the spectrum, it is necessary to consider the interaction that protons and other atomic nuclei have experienced in the background light field~\footnote{Besides CMB photons which represent the densest photon background, protons and nuclei interact mainly with infrared, optical and ultra-violet photons~(see a review see Ref.~\cite{P54-Allard-2011-extragalactic}).}.
Besides the adiabatic losses due to the expansion of the Universe, protons mainly suﬀer from the electron-positron pair production mechanism, whose energy threshold with CMB photons is around $10^{18}~\mathrm{eV}$~(the early work was carried out by Hillas~\cite{P46-Hillas-1967-the} and Blumenthal~\cite{P47-Blumenthal-1970-energy}), and pion production, which is the commonly known GZK mechanism.
The interactions experienced by nuclei with photon backgrounds are different from the protons~(for a review see Ref.~\cite{P48-Rachen-1996-interaction}).
There are three categories of nucleus energy loss processes: the adiabatic expansion, the pair production mechanism and the nucleus photodisintegration process. 
For the photodisintegration process, the energy threshold is proportional to the element mass, and diﬀerent processes become dominant at diﬀerent energies: the dominant processes with energy from ground to high are giant dipole resonance~\footnote{The giant dipole resonance is the lowest energy and highest cross section process, which is the collective excitation of the nucleus in response to electromagnetic radiation between $10\sim50~\mathrm{MeV}$.}, quasi-deuteron process~\footnote{Around $30~\mathrm{MeV}$ in the nucleus rest frame, the quasi-deuteron process becomes comparable to the giant dipole resonance and its contribution dominates the total cross section at higher energies.} and photopion production process~\footnote{The photopion production~(or baryonic resonances) of nuclei becomes relevant above $150~\mathrm{MeV}$ in the nuclei rest frame (e.g., $\sim 5 \times 10^{21}~\mathrm{eV}$ in the lab frame for iron nuclei interacting with the CMB).}.
As a result of the very fast photodisintegration with CMB photons, elemental groups should simply disappear from the spectrum at high energy one after the other~ (see for instance~\cite{P49-Allard-2008-implications} for a more complete discussion).
High resolution measurements of the composition have been allowed by recent experiments like HiRes~\cite{P50-HiRes-2009-indications}, the Pierre Auger Observatory~\cite{P51-PierreAuger-2010-measurement,P52-PierreAuger-2014-depth}, and Telescope Array~\cite{P53-TelescopeArray-2018-depth}.
However, there is no unified opinion on the quality composition of the GZK region: HiRes and Telescope Array results are compatible with light primary scenario resembling mostly protons, while the Auger results shows an indication towards heavier composition.
From the results of HiRes and Telescope Array, our assumption, that protons dominate the cosmic-ray structure in the GZK region, is valid, while the results of Pierre Auger Observatory, that indicate an evolution of the composition toward heavier elements, mean our assumption needs to be used with caution. \\

From the above discussion, we see that the current GZK region observations have set strict constraints on the Case III parameter space. 
However, for the Case I and II regions, further experimental observations and theoretical researches are needed.
From the observational perspective, the sources and energies of UHE cosmic rays need to be further investigated.
In classic case, the proton mean free path in the CMB photons decreases exponentially with energy~(down to a few $\mathrm{Mpc}$) above the GZK limit.
In LV case, the accurate detection of the cosmic-ray source distances contributes to further study on the LV effects of proton and pion.
In theoretical researches, the further considering of the proton and photon reaction channel is useful, such as $p+\gamma\to\Delta(1232)$.\\

\section{Conclusion}

Using the LV modification framework of the proton and pion dispersion relations, we get different threshold behaviors of pion production off the proton.
We obtain the corresponding constraints on proton and pion LV parameters for linear, quadratic, and $n$-order cases.
Besides the conventional GZK cut-off, we also predict abnormal threshold behaviors under different LV situations, in similar to the photon case as discussed in the literature~\cite{P41-Li-2021-threshold,P42-Li-2021-ultrahigh,P43-Li-2022-searching,P44-Li-2022-lorentz,P45-Li-2023-revisiting}.
The LV parameter space has been sorted into three regions.
For Case I, the threshold behavior is similar to the classic case, and the LV effect is difficult to be detected from spectrum observations.
In Case II, there are two thresholds for the proton collision reaction, and we might observe the reappearance of extremely high energy protons.
This cosmic ray spectrum structure need further observation to testify its existence.
In Case III, the LV effects are obvious, and current GZK region observations have set strict constraints on the space of LV parameters. 
Although previous experimental observations have set strict constraints on the LV effect, it actually only means a strong constraint on the LV effect in Case III. 
For the Cases I and II, there is still space to explore the quantum gravity and grand unified theory that allow for some specific LV effect.
Further experimental observations and theoretical researches are needed to constrain the LV effects around GZK regions. \\

This work is supported by National Natural Science Foundation of China (Grants No.~12335006 and No.~12075003).


\begin{thebibliography}{99}

\bibitem{P36-Linsley-1961-extremely}
J.~Linsley, L.~Scarsi and B.~Rossi,
``{Extremely energetic cosmic-ray event},''
\href{https://doi.org/10.1103/PhysRevLett.6.485}{Phys. Rev. Lett. {\bf 6}, 485-487 (1961)}.

\bibitem{P37-Linsley-1963-evidence}
J.~Linsley,
``{Evidence for a primary cosmic-ray particle with energy $10^{20}~\mathrm{eV}$},''
\href{https://doi.org/10.1103/PhysRevLett.10.146}{Phys. Rev. Lett. {\bf 10}, 146-148 (1963)}.

\bibitem{P1-Greisen-1966-end}
K.~Greisen,
``{End to the cosmic-ray spectrum?},''
\href{https://doi.org/10.1103/PhysRevLett.16.748}{Phys. Rev. Lett. {\bf 16}, 748-750 (1966)}. 

\bibitem{P2-Zatsepin-1966-upper}
G.~T.~Zatsepin and V.~A.~Kuzmin,
``{Upper limit of the spectrum of cosmic rays},''
{JETP Lett. {\bf 4}, 78-80 (1966)}. 

\bibitem{P3-Sokolsky-2007-Highest}
P.~Sokolsky and G.~B.~Thomson,
``{Highest energy cosmic-rays and results from the HiRes experiment},''
\href{https://doi.org/10.1088/0954-3899/34/11/R01}{J. Phys. G {\bf 34}, R401 (2007)} 
[\href{https://arxiv.org/abs/0706.1248}{\tt arXiv:0706.1248}].

\bibitem{P4-Bird-1994-detection}
D.~J.~Bird \textit{et al.},
``{Detection of a cosmic ray with measured energy well beyond the expected spectral cutoff due to cosmic microwave radiation},''
\href{https://doi.org/10.1086/175344}{Astrophys. J. {\bf 441}, 144-150 (1995)} 
[\href{https://arxiv.org/abs/astro-ph/9410067}{\tt arXiv:astro-ph/9410067}].

\bibitem{P5-Takeda-1998-extension}
M.~Takeda \textit{et al.},
``{Extension of the cosmic-ray energy spectrum beyond the predicted Greisen-Zatsepin-Kuz'min cutoff},''
\href{https://doi.org/10.1103/PhysRevLett.81.1163}{Phys. Rev. Lett. {\bf 81}, 1163-1166 (1998)} 
[\href{https://arxiv.org/abs/astro-ph/9807193}{\tt arXiv:astro-ph/9807193}].

\bibitem{P13-Kephart-1995-magnrtic}
T.~W.~Kephart and T.~J.~Weiler,
``{Magnetic monopoles as the highest energy cosmic ray primaries},''
\href{https://doi.org/10.1016/0927-6505(95)00043-7}{Astropart. Phys. {\bf 4}, 271-279 (1996)} 
[\href{https://arxiv.org/abs/astro-ph/9505134}{\tt arXiv:astro-ph/9505134}].

\bibitem{P14-Weiler-1997-cosmic}
T.~J.~Weiler,
``{Cosmic-ray neutrino annihilation on relic neutrinos revisited: a mechanism for generating air showers above the Greisen-Zatsepin-Kuzmin cutoff},''
\href{https://doi.org/10.1016/S0927-6505(98)00068-1}{Astropart. Phys. {\bf 11}, 303-316 (1999)} 
[\href{https://arxiv.org/abs/hep-ph/9710431}{\tt arXiv:hep-ph/9710431}].

\bibitem{P15-Berezinsky-1997-ultrahigh}
V.~Berezinsky, M.~Kachelriess and A.~Vilenkin,
``{Ultrahigh energy cosmic rays without GZK cutoff},''
\href{https://doi.org/10.1103/PhysRevLett.79.4302}{Phys. Rev. Lett. {\bf 79}, 4302-4305 (1997)} 
[\href{https://arxiv.org/abs/astro-ph/9708217}{\tt arXiv:astro-ph/9708217}].

\bibitem{P12-Farrar-1999-GZK}
G.~R.~Farrar and T.~Piran,
``{Violation of the Greisen-Zatsepin-Kuzmin cutoff: a tempest in a (magnetic) teapot? Why cosmic ray energies above $10^{20}~\mathrm{eV}$ may not require new physics},''
\href{https://doi.org/10.1103/PhysRevLett.84.3527}{Phys. Rev. Lett. {\bf 84}, 3527 (2000)} 
[\href{https://arxiv.org/abs/astro-ph/9906431}{\tt arXiv:astro-ph/9906431}].

\bibitem{P20-Coleman-1998-high}
S.~R.~Coleman and S.~L.~Glashow,
``{High-energy tests of Lorentz invariance},''
\href{https://doi.org/10.1103/PhysRevD.59.116008}{Phys. Rev. D {\bf 59}, 116008 (1999)} 
[\href{https://arxiv.org/abs/hep-ph/9812418}{\tt arXiv:hep-ph/9812418}].

\bibitem{P21-Coleman-1998-evading}
S.~R.~Coleman and S.~L.~Glashow,
``{Evading the GZK cosmic-ray cutoff},''
[\href{https://arxiv.org/abs/hep-ph/9808446}{\tt arXiv:hep-ph/9808446}].

\bibitem{P22-Amelino-2000-planck}
G.~Amelino-Camelia and T.~Piran,
``{Planck-scale deformation of Lorentz symmetry as a solution to the ultrahigh energy cosmic ray and the TeV-photon paradoxes},''
\href{https://doi.org/10.1103/PhysRevD.64.036005}{Phys. Rev. D {\bf 64}, 036005 (2001)} 
[\href{https://arxiv.org/abs/astro-ph/0008107}{\tt arXiv:astro-ph/0008107}].

\bibitem{P23-Amelino-2001-space}
G.~Amelino-Camelia,
``{Space-time quantum solves three experimental paradoxes},''
\href{https://doi.org/10.1016/S0370-2693(02)01223-6}{Phys. Lett. B {\bf 528}, 181-187 (2002)} 
[\href{https://arxiv.org/abs/gr-qc/0107086}{\tt arXiv:gr-qc/0107086}].

\bibitem{P25-Gonzalez-1997-vacuum}
L.~Gonzalez-Mestres,
``{Vacuum structure, Lorentz symmetry and superluminal particles},''
[\href{https://arxiv.org/abs/physics/9704017}{\tt arXiv:physics/9704017}].

\bibitem{P26-Sato-2000-extremely}
H.~Sato,
``{Extremely high-energy and violation of Lorentz invariance},''
[\href{https://arxiv.org/abs/astro-ph/0005218}{\tt arXiv:astro-ph/0005218}].

\bibitem{P27-Bertolami-1999-proposed}
O.~Bertolami and C.~S.~Carvalho,
``{Proposed astrophysical test of Lorentz invariance},''
\href{https://doi.org/10.1103/PhysRevD.61.103002}{Phys. Rev. D {\bf 61}, 103002 (2000)} 
[\href{https://arxiv.org/abs/gr-qc/9912117}{\tt arXiv:gr-qc/9912117}].

\bibitem{P28-Aloisio-2000-probing}
R.~Aloisio, P.~Blasi, P.~L.~Ghia and A.~F.~Grillo,
``{Probing the structure of space-time with cosmic rays},''
\href{https://doi.org/10.1103/PhysRevD.62.053010}{Phys. Rev. D {\bf 62}, 053010 (2000)} 
[\href{https://arxiv.org/abs/astro-ph/0001258}{\tt arXiv:astro-ph/0001258}].

\bibitem{P7-Thomson-2004-new}
G.~Thomson,
``{New results from the HiRes experiment},''
\href{https://doi.org/10.1016/j.nuclphysbps.2004.10.061}{Nucl. Phys. B Proc. Suppl. {\bf 136}, 28-33 (2004)}. 

\bibitem{P8-HiRes-2004-measyrement}
R.~U.~Abbasi \textit{et al.} [HiRes],
``{Measurement of the flux of ultrahigh energy cosmic rays from monocular observations by the High Resolution Fly's Eye experiment},''
\href{https://doi.org/10.1103/PhysRevLett.92.151101}{Phys. Rev. Lett. {\bf 92}, 151101 (2004)} 
[\href{https://arxiv.org/abs/astro-ph/0208243}{\tt arXiv:astro-ph/0208243}].

\bibitem{P9-HiRes-2005-observation}
R.~U.~Abbasi \textit{et al.} [HiRes],
``{Observation of the ankle and evidence for a high-energy break in the cosmic ray spectrum},''
\href{https://doi.org/10.1016/j.physletb.2005.05.064}{Phys. Lett. B {\bf 619}, 271-280 (2005)} 
[\href{https://arxiv.org/abs/astro-ph/0501317}{\tt arXiv:astro-ph/0501317}].

\bibitem{P10-HiRes-2008-first}
R.~U.~Abbasi \textit{et al.} [HiRes],
``{First observation of the Greisen-Zatsepin-Kuzmin suppression},''
\href{https://doi.org/10.1103/PhysRevLett.100.101101}{Phys. Rev. Lett. {\bf 100}, 101101 (2008)} 
[\href{https://arxiv.org/abs/astro-ph/0703099}{\tt arXiv:astro-ph/0703099}].

\bibitem{P29-PierreAuger-2020-features}
A.~Aab \textit{et al.} [Pierre Auger],
``{Features of the energy spectrum of cosmic rays above $2.5\times10^{18}~\mathrm{eV}$ using the Pierre Auger Observatory},''
\href{https://doi.org/10.1103/PhysRevLett.125.121106}{Phys. Rev. Lett. {\bf 125}, 121106 (2020)} 
[\href{https://arxiv.org/abs/2008.06488}{\tt arXiv:2008.06488}].

\bibitem{P30-PierreAuger-2020-measurement}
A.~Aab \textit{et al.} [Pierre Auger],
``{Measurement of the cosmic-ray energy spectrum above $2.5\times10^{18}~\mathrm{eV}$ using the Pierre Auger Observatory},''
\href{https://doi.org/10.1103/PhysRevD.102.062005}{Phys. Rev. D {\bf 102}, 062005 (2020)} 
[\href{https://arxiv.org/abs/2008.06486}{\tt arXiv:2008.06486}].

\bibitem{P35-Abbasi-2023-the}
R.~U.~Abbasi \textit{et al.},
``{The energy spectrum of cosmic rays measured by the Telescope Array using 10 years of fluorescence detector data},''
\href{https://doi.org/10.1016/j.astropartphys.2023.102864}{Astropart. Phys. {\bf 151}, 102864 (2023)}.

\bibitem{P38-Xiao-2008-lorentz}
Z.~Xiao and B.-Q.~Ma,
``{Lorentz violation dispersion relation and its application},''
\href{https://doi.org/10.1142/S0217751X09042955}{Int. J. Mod. Phys. A {\bf 24}, 1359-1381 (2009)} 
[\href{https://arxiv.org/abs/0805.2012}{\tt arXiv:0805.2012}].

\bibitem{P39-Bi-2008-testing}
X.~J.~Bi, Z.~Cao, Y.~Li and Q.~Yuan,
``{Testing Lorentz invariance with ultrahigh energy cosmic ray spectrum},''
\href{https://doi.org/10.1103/PhysRevD.79.083015}{Phys. Rev. D {\bf 79}, 083015 (2009)} 
[\href{https://arxiv.org/abs/0812.0121}{\tt arXiv:0812.0121}].

\bibitem{P40-Stecker-2009-searching}
F.~W.~Stecker and S.~T.~Scully,
``{Searching for new physics with ultrahigh energy cosmic rays},''
\href{https://doi.org/10.1088/1367-2630/11/8/085003}{New J. Phys. {\bf 11}, 085003 (2009)} 
[\href{https://arxiv.org/abs/0906.1735}{\tt arXiv:0906.1735}].

\bibitem{P41-Li-2021-threshold}
H.~Li and B.-Q.~Ma,
``{Threshold anomalies of ultra-high energy cosmic photons due to Lorentz invariance violation},''
\href{https://doi.org/10.1016/j.jheap.2021.07.001}{JHEAp {\bf 32}, 1-5 (2021)} 
[\href{https://arxiv.org/abs/2105.06647}{\tt arXiv:2105.06647}].

\bibitem{P42-Li-2021-ultrahigh}
C.~Li and B.-Q.~Ma,
``{Ultrahigh-energy photons from LHAASO as probes of Lorentz symmetry violations},''
\href{https://doi.org/10.1103/PhysRevD.104.063012}{Phys. Rev. D {\bf 104}, no.6, 063012 (2021)} 
[\href{https://arxiv.org/abs/2105.07967}{\tt arXiv:2105.07967}].

\bibitem{P43-Li-2022-searching}
H.~Li and B.-Q.~Ma,
``{Searching Lorentz invariance violation from cosmic photon attenuation},''
\href{https://doi.org/10.1140/epjc/s10052-023-11334-z}{Eur. Phys. J. C {\bf 83}, no.3, 192 (2023)} 
[\href{https://arxiv.org/abs/2210.05563}{\tt arXiv:2210.05563}].

\bibitem{P44-Li-2022-lorentz}
H.~Li and B.-Q.~Ma,
``{Lorentz invariance violation induced threshold anomaly versus very-high energy cosmic photon emission from GRB 221009A},''
\href{https://doi.org/10.1016/j.astropartphys.2023.102831}{Astropart. Phys. {\bf 148}, 102831 (2023)} 
[\href{https://arxiv.org/abs/2210.06338}{\tt arXiv:2210.06338}].

\bibitem{P45-Li-2023-revisiting}
H.~Li and B-Q.~Ma,
``{Revisiting Lorentz invariance violation from GRB 221009A},'' 
\href{ https://doi.org/10.1088/1475-7516/2023/10/061}{JCAP {\bf 10}, 061 (2023)}
[\href{https://arxiv.org/abs/2306.02962}{\tt arXiv:2306.02962}].

\bibitem{P57-PierreAuger-2021-testing}
P.~Abreu \textit{et al.} [Pierre Auger],
``{Testing effects of Lorentz invariance violation in the propagation of astroparticles with the Pierre Auger Observatory},'' 
\href{https://doi.org/10.1088/1475-7516/2022/01/023}{JCAP {\bf 01}, no.01, 023 (2022)}
[\href{https://arxiv.org/abs/2112.06773}{\tt arXiv:2112.06773}].

\bibitem{P58-Maccione-2009-Planck}
L.~Maccione, A.~M.~Taylor, D.~M.~Mattingly and S.~Liberati,
``{Planck-scale Lorentz violation constrained by Ultra-High-Energy Cosmic Rays},'' 
\href{https://doi.org/10.1088/1475-7516/2009/04/022}{JCAP {\bf 04}, 022 (2009)}
[\href{https://arxiv.org/abs/0902.1756}{\tt arXiv:0902.1756}].

\bibitem{P18-H6-Mattingly-2002-threshold}
D.~Mattingly, T.~Jacobson and S.~Liberati,
``{Threshold configurations in the presence of Lorentz violating dispersion relations},''
\href{https://doi.org/10.1103/PhysRevD.67.124012}{Phys. Rev. D {\bf 67}, 124012 (2003)} 
[\href{https://arxiv.org/abs/hep-ph/0211466}{\tt arXiv:hep-ph/0211466}].

\bibitem{P19-H18-He-2022-lorentz}
P.~He and B.-Q.~Ma,
``{Lorentz symmetry violation of cosmic photons},''
\href{https://doi.org/10.3390/universe8060323}{Universe {\bf 8}, no.6, 323 (2022)} 
[\href{https://arxiv.org/abs/2206.08180}{\tt arXiv:2206.08180}].

\bibitem{P55-Lang-2020-Testing}
R.~G.~Lang [Pierre Auger],
``{Testing Lorentz Invariance Violation at the Pierre Auger Observatory},''
\href{https://doi.org/10.22323/1.358.0327}{PoS {\bf ICRC2019}, 327 (2020)}.

%\bibitem{P56-Abreu-2021-Testing}
%P.~Abreu \textit{et al.} [Pierre Auger],
%``{Testing effects of Lorentz invariance violation in the propagation of astroparticles with the Pierre Auger Observatory},''
%\href{https://doi.org/10.1088/1475-7516/2022/01/023}{JCAP {\bf 01}, no.01, 023 (2022)} 
%[\href{https://arxiv.org/abs/2112.06773}{\tt arXiv:2112.06773}].

\bibitem{P59-Jacobson-2002-Threshold}
T.~Jacobson, S.~Liberati and D.~Mattingly,
``{Threshold effects and Planck scale Lorentz violation: Combined constraints from high-energy astrophysics},''
\href{https://doi.org/10.1103/PhysRevD.67.124011}{Phys. Rev. D {\bf 67}, 124011 (2003)} 
[\href{https://arxiv.org/abs/hep-ph/0209264}{\tt arXiv:hep-ph/0209264}].

\bibitem{P54-Allard-2011-extragalactic}
D.~Allard,
``{Extragalactic propagation of ultrahigh energy cosmic-rays},''
\href{https://doi.org/10.1016/j.astropartphys.2011.10.011}{Astropart. Phys. {\bf 39-40}, 33-43 (2012)} 
[\href{https://arxiv.org/abs/1111.3290}{\tt arXiv:1111.3290}].

\bibitem{P46-Hillas-1967-the}
A.M.~Hillas,
``{The energy spectrum of cosmic rays in an evolving universe},'' 
\href{https://doi.org/10.1016/0375-9601(67)91023-7}{Phys. Lett. A {\bf 24} 677 (1967)}.

\bibitem{P47-Blumenthal-1970-energy}
G.~R.~Blumenthal,
``{Energy loss of high-energy cosmic rays in pair-producing collisions with ambient photons},''
\href{https://doi.org/10.1103/PhysRevD.1.1596}{Phys. Rev. D {\bf 1} 1596 (1970)}.

\bibitem{P48-Rachen-1996-interaction}
J.~P.~Rachen,
``{Interaction Processes and Statistical Properties of the Propagation of Cosmic Rays in Photon Backgrounds},''
\href{ https://doi.org/10.5281/zenodo.3242300}{PhD thesis of the Bohn University}.

\bibitem{P49-Allard-2008-implications}
D.~Allard, N.~G.~Busca, G.~Decerprit, A.~V.~Olinto and E.~Parizot,
``{Implications of the cosmic ray spectrum for the mass composition at the highest energies},''
\href{https://doi.org/10.1088/1475-7516/2008/10/033}{JCAP {\bf 10}, 033 (2008)} 
[\href{https://arxiv.org/abs/0805.4779}{\tt arXiv:0805.4779}].

\bibitem{P50-HiRes-2009-indications}
R.~U.~Abbasi \textit{et al.} [HiRes],
``{Indications of Proton-Dominated Cosmic Ray Composition above $1.6~\mathrm{EeV}$},''
\href{https://doi.org/10.1103/PhysRevLett.104.161101}{Phys. Rev. Lett. {\bf 104}, 161101 (2010)} 
[\href{https://arxiv.org/abs/0910.4184}{\tt arXiv:0910.4184}].

\bibitem{P51-PierreAuger-2010-measurement}
J.~Abraham \textit{et al.} [Pierre Auger],
``{Measurement of the Depth of Maximum of Extensive Air Showers above $10^{18}~\mathrm{eV}$},''
\href{https://doi.org/10.1103/PhysRevLett.104.091101}{Phys. Rev. Lett. {\bf 104}, 091101 (2010)} 
[\href{https://arxiv.org/abs/1002.0699}{\tt arXiv:1002.0699}].

\bibitem{P52-PierreAuger-2014-depth}
A.~Aab \textit{et al.} [Pierre Auger],
``Depth of maximum of air-shower profiles at the Pierre Auger Observatory. II. Composition implications,''
\href{https://doi.org/10.1103/PhysRevD.90.122006}{Phys. Rev. D {\bf 90}, no.12, 122006 (2014)} 
[\href{https://arxiv.org/abs/1409.5083}{\tt arXiv:1409.5083}].

\bibitem{P53-TelescopeArray-2018-depth}
R.~U.~Abbasi \textit{et al.} [Telescope Array],
``Depth of Ultra High Energy Cosmic Ray Induced Air Shower Maxima Measured by the Telescope Array Black Rock and Long Ridge FADC Fluorescence Detectors and Surface Array in Hybrid Mode,''
\href{https://doi.org/10.3847/1538-4357/aabad7}{Astrophys. J. {\bf 858}, no.2, 76 (2018)} 
[\href{https://arxiv.org/abs/1801.09784}{\tt arXiv:1801.09784}].



\end{thebibliography}
\end{document}